\documentstyle[preprint,aps]{revtex}
\newcommand{\be}{\begin{equation}} \newcommand{\ee}{\end{equation}} 
\newcommand{\bea}{\begin{eqnarray}}\newcommand{\eea}{\end{eqnarray}}

\begin{document}
\draft
\preprint{MRI-PHY/97-01, hep-th/9701132}
\title { Solitons in 1+1 Dimensional Gauged Sigma Models}
\author{Pijush K. Ghosh$^{*}$}
\address{The Mehta Research Institute of
Mathematics \& Mathematical Physics,\\
Chhatnag Road, Jhusi, Allahabad-221506, U.P., INDIA.}
\footnotetext {$\mbox{}^*$ E-mail: 
pijush@mri.ernet.in }  
\maketitle
\begin{abstract} 
We study soliton solutions in 1+1 dimensional gauged sigma models,
obtained by dimensional reduction from its 2+1 dimensional counterparts.
We show that the Bogomol'nyi bound of these models can be expressed in
terms of two conserved charges in a similar way to that of
the BPS dyons in $3+1$ dimensions.
Purely magnetic vortices of the 2+1 dimensional completely gauged 
sigma model
appear as charged solitons in the corresponding 1+1 dimensional theory. 
The scale invariance of these solitons is also
broken because of the dimensional reduction. We
obtain exact static soliton solutions of these models saturating the 
Bogomol'nyi bound. 
\end{abstract}
%keyword(s): sigma model; Bogomol'nyi bound; dyons; vortices.}
\narrowtext

\newpage
\section{Introduction}

Recently, there has been much interest in the study of soliton solutions
in 2+1 dimensional gauged sigma models \cite{nardeli,g1,dur,dur1,gg,g2,may}.
These models can be viewed as a low energy effective action of certain
gauged linear sigma models with judiciously chosen Higgs potential. 
The self-dual soliton solutions
of the completely gauged sigma model with pure Chern-Simons(CS) dynamics are
scale invariant \cite{nardeli}. In fact, the scalar multiplet is exactly
equivalent to that of the usual sigma model \cite{sig}.
Moreover, purely magnetic
vortices can be obtained from this completely gauged sigma model after a
suitable reduction of the non-Abelian gauge group, as in the case of
't Hooft-Polyakov monopole \cite{mono}. Unlike in the monopole case,
these magnetic vortices can also be obtained from an Abelian theory,
since the Bianchi identity for the $U(1)$ invariant field strength
is satisfied in $2+1$ dimensions. All these results can be generalized
to the completely gauged $CP^N$ models, either with pure CS \cite{nardeli}
or
Yang-Mills CS dynamics \cite{g1}.

Solitons of the 2+1 dimensional $U(1)$ gauged
sigma models are not scale invariant, because
of the presence of a $U(1)$ invariant potential term \cite{dur,gg}.
No exact solution is known for any of these models.
However, it is known that these models admit a variety of new
soliton solutions \cite{dur,gg,may}. One such peculiar feature is that
the magnetic flux of the topological solitons
in the symmetric phase of the theory is not necessarily quantized like
vortices in the Abelian Higgs models. The quantization of the magnetic flux
is recovered in
the asymmetric phase. These sigma models with pure CS dynamics admit both
topological as well as nontopological soliton solutions, as in the case of
self-dual $U(1)$ CS Higgs theory. In fact,
these $U(1)$ gauged sigma models can be reduced to the 
self-dual $U(1)$ CS Higgs theory \cite{sd} in certain limits. All these
results have been
generalized to the gauged $CP^N$ case with pure CS dynamics \cite{g2},
where a proper subgroup of the global $SU(N+1)$ is completely gauged. 

The 1+1 dimensional sigma models have many properties in common
with the 3+1 dimensional Yang-Mills-Higgs(YMH) theory. One such remarkable
feature
is that a class of $1+1$ dimensional sigma models admit Q-kinks \cite{cam,rl}
with
similar properties to those of the BPS dyons of YMH theory \cite{bps}.
These
1+1 dimensional sigma models are obtained by dimensional reduction from their
2+1 dimensional counterparts. It is reasonable at this point to
study dimensionally reduced version of the $2+1$ dimensional gauged sigma
models.

The purpose of this paper is to study soliton solutions in 1+1 dimensional
gauged sigma models obtained by dimensional reduction from 2+1 dimensions.
In particular, we consider dimensionally reduced version
of two different 2+1 dimensional models with pure CS dynamics,
(i) the completely gauged sigma model \cite{nardeli} 
and (ii) the $U(1)$ gauged sigma model \cite{gg}.
We find that the Bogomol'nyi bound \cite{bogo} for both of these models are
of BPS type,
namely, the lower bound on the energy is expressed as a linear combination
of the topological charge and the Noether charge. Moreover, the scale
invariant soliton solutions, describing purely magnetic
vortices, of the 2+1 dimensional completely gauged sigma model, appear as
solitons solutions with nonzero Noether charge and a definite scale,
in the corresponding 1+1 dimensional theory. 
This resembles the
way BPS dyons in 3+1 dimensions can be obtained from four 
dimensional Eucledian
self-dual Yang-Mills theory. Such an analog already exists in the framework
of certain 1+1 dimensional sigma models admitting Q-kinks \cite{cam} which
are necessarily time dependent solutions. However, in our case the similarity
is between static solitons of the completely gauged model and 3+1 dimensional
dyons. Recently, it has been shown that the dimensionally reduced version
of self-dual $U(1)$ CS Higgs theory also shares similar
properties \cite{kor}. Unfortunately, 
the soliton solutions of 2+1 dimensional $U(1)$ CS Higgs theory are not
scale invariant. Thus, the completely gauged sigma model studied in this
paper have more similarities with BPS dyons than any other existing models. 
We find all the static,
exact soliton solutions of the completely gauged sigma model saturating
the Bogomol'nyi bound.
We are able to obtain only a class of exact soliton solutions for the
$U(1)$ gauged sigma model. The soliton solutions of both of these models
are domain walls, interpolating between different symmetric and asymmetric
vacua.

The plan of this paper is the following. First, we introduce and study the
1+1 dimensional completely gauged sigma model in Sec. II. In particular,
we describe the dimensional reduction procedure and obtain the 1+1
dimensional model from the 2+1 dimensional completely gauged sigma model.
We obtain the Bogomol'nyi bound of this model and present all static, exact
solutions of the Bogomol'nyi equations. In Sec. III,
we discuss
the $U(1)$ gauged sigma model and obtain a set of exact, static soliton
solutions saturating the Bogomol'nyi bound. Finally, we give a summary of the
results obtained in this paper and 
discuss on possible directions to be explored in Sec. IV. In appendix
A, we present some more exact soliton solutions saturating the Bogomol'nyi
bound for the $U(1)$ gauged sigma model.

\section{$O(3)$ Gauged Sigma Model}

The self-dual completely gauged sigma model in 2+1 dimensions is given by
\cite{nardeli},
\be
{\cal{L}}_0 = \frac{1}{2} D_\mu \chi^a D^\mu \chi^a + \frac{\kappa}{4}
\epsilon^{\mu \nu \lambda} \left ( F_{\mu \nu}^a A_\lambda^a -
\frac{e}{3} \epsilon^{a b c} A_\mu^a A_\nu^b A_\lambda^c \right ) .
\label{e.0}
\ee
\noindent The real scalar field $\chi$ has three components and is
constrained to lie on a unit sphere in the
internal space, i.e., $\chi^a \chi^a =1$. The covariant derivative and
the field strengths $F_{01}^a$ are defined as,
\be
D_\mu \chi^a = \partial_\mu \chi^a + e \epsilon^{a b c} A_\mu^b \chi^c, \ \
F_{\mu \nu}^a = \partial_\mu A_\nu^a - \partial_\nu A_\mu^a +
e \epsilon^{a b c} A_\mu^b A_\nu^c . 
\label{eq4.1}
\ee
\noindent The self-dual equations of (\ref{e.0}) are,
\be
D_0 \chi^a = 0, \ \ D_i \chi^a \pm \epsilon_{ij} \epsilon^{a b c}
\chi^b D_j \chi^c =0 .
\label{eq4.1.1}
\ee
\noindent The self-dual soliton solutions of (\ref{e.0}) are 
characterized by 
the energy $E= 4 \pi q_0$, where the topological charge $q_0$ is defined as,
\be
q_0= \int d^2 x k_0, \ \ k_\mu = \frac{1}{2} \epsilon_{\mu \nu \lambda}
{\cal{F}}^{\nu \lambda}, \ \ {\cal{F}}_{\mu \nu} =
\epsilon^{a b c} D_\mu \chi^a D_\nu \chi^b \chi^c - e F_{\mu \nu}^a \chi^a.
\label{eq4.2.2}
\ee
\noindent Here, $k_\mu$ is the gauge invariant topological current and
${\cal{F}}_{\mu \nu}$ is the $U(1)$ invariant field strength.
The conservation of the topological current automatically implies
the Bianchi identity for ${\cal{F}}_{\mu \nu}$. Note that the
electric field ${\cal{F}}_{01}$ vanishes for the self-dual field
configurations, since the Gauss law implies
$F_{\mu \nu}^a \chi^a =0$. However, the magnetic field is nonzero
and it describes Liouville vortex. These self-dual field configurations
are of zero Noether charge. As a consequence, the spatial components of
the gauge
fields are pure gauges and the second equation of (\ref{eq4.2.2}) can be
exactly mapped into the corresponding equation of the usual sigma model.
Thus, the solitons are scale invariant.

We now dimensionally reduce the model (\ref{e.0}) to
1+1 dimensions. We take all the field variables to be independent
of the second coordinate and identify the second component of the
gauge field with a triplet $M^a$, $A_2^a=M^a$. Following the standard
procedure, we have the $1+1$ dimensional completely gauged
sigma model,
\be
{\cal{L}}_1 = \frac{1}{2} ( D_\mu \chi^a )^2 - e^2 \left ( M^a M^a -
( M^a \chi^a )^2 \right ) + \kappa M^a F_{01}^a, \ \ \mu=0, 1.
\label{eq4.0}
\ee
\noindent Note the appearance of an interaction term. 
The vacuum is characterized by those field configurations, for which
the norm of the triplet $M$ is exactly equal to the square of its own
projection along $\chi$.

The equations of motion of (\ref{eq4.0}) are,
\bea
& & \kappa F_{01}^a = e^2 \left [  M^a - \chi^a (M^b \chi^b) \right ], \ \ \
\kappa D_0 M^a = K_1^a, \ \ \ \kappa D_1 M^a = K_0^a,\nonumber \\
& & D_\mu K^{a,\mu} = e^3 \epsilon^{a b c} M^b \chi^c ( M^d \chi^d),
\ \ \ K_\mu^a = e \epsilon^{a b c} D_\mu \chi^b \ \chi^c, 
\label{eq4.2}
\eea
\noindent where a prime denotes differentiation with respect to the
space coordinate. 
Note that $F_{01}^a \chi^a=0$. Also, $ \chi^a D_\mu M^a =0$,
since $K_\mu^a \chi^a=0$.

We now define two conserved charges $Y_1$ and $Z_1$,
\be
Y_1 = \kappa \int dx \left [ M^a M^a \right]^\prime, \ \ \
Z_1 =2 e \int dx \left [ \chi^a M^a \right ]^\prime .
\label{eq4.3a}
\ee
\noindent These two charges, $Y_1$ and $Z_1$, can be identified as 
topological and Noether charges respectively. The reason for such an
identification is the following. The momentum along the compactified
direction, $P_y$, is a conserved quantity in the $2+1$ dimensional theory,
because of the translational
invariance. This quantity
remains conserved in the corresponding $1+1$ dimensional model also.
Using the Gauss law, we find,
\be
P_y = - \frac{\kappa}{2} \int dx \partial_1 \left ( M^a M^a \right ).
\label{eq4.3b}
\ee
\noindent Thus, $Y_1$ can be identified as the topological charge.
In order to identify $Z_1$ as the Noether charge, recall that we are
dealing with pure CS dynamics. Consequently, the topological current $k_\mu$
, as defined in the second equation of (\ref{eq4.2.2}), can be identified as the
$U(1)$ current in the asymmetric phase of the $2+1$ dimensional theory. This
$U(1)$ charge remains conserved
in the corresponding $1+1$ dimensional theory also. In particular, the
topological charge $k_0$ ( or equivalently the $U(1)$ Noether charge ) 
reduces to the
following expression after the dimensional reduction,
\be
{\cal{F}}_{12} =- e D_1 \chi^a M^a = -e \partial_1 \left ( \chi^a M^a \right ).
\label{eq4.3c}
\ee
%\noindent after the compactification. 
%Note that we are dealing with pure CS
%dynamics and hence, $\int dx {\cal{F}}_{12}$ can be
%identified as the total $U(1)$ charge in the asymmetric phase.
\noindent Thus, $Z_1$ is identified as the Noether charge.

The energy functional can be written as,
\bea
E & = & \int dx \left [ \left ( D_0 \chi^a \pm e \cos{\beta}
\epsilon^{a b c} M^b \chi^c \right )^2
+ \left ( D_1 \chi^a \mp e \sin{\beta} \left \{ M^a -
\chi^a \left ( \chi^b M^b \right ) 
\right \} \right )^2 \right ]\nonumber \\
& & \pm ( Y_1 \cos{\beta} + Z_1 \sin{\beta} ) . 
\label{4.3}
\eea
\noindent The lower bound on the energy functional is saturated for the 
solutions of the following first order equations,
\be
D_0 \chi^a \pm e \cos{\beta} \epsilon^{a b c} M^b
\chi^c =0, \ \ \
D_1 \chi^a \mp e \sin{\beta} \left [ M^a - \chi^a \left ( \chi^b M^b \right ) 
\right ] =0.
\label{eq4.4}
\ee
\noindent These two first order equations are consistent with the
second order field equations. The gauge potential $A_0^a$ is
determined in terms of $\chi^a$ and $M^a$
from the first equation of
(\ref{eq4.2}) and (\ref{eq4.4}) respectively as,
\be
A_0^a = \mp \frac{e}{\kappa} \sin{\beta} \chi^a \mp \cos{\beta} M^a.
\label{eq4.5}
\ee 
\noindent Using the Gauss law and the first Bogomol'nyi
equation of (\ref{eq4.4}), we have,
\be
D_1 M^a \mp \frac{e^2}{\kappa} \cos{\beta} \left [ M^a - \chi^a (\chi^b
M^b) \right ] = 0.
\label{eq4.6}
\ee
\noindent Multiplying the second equation of (\ref{eq4.4}) and Eq.
(\ref{eq4.6}), respectively  by $M^a$, we find,
\bea
&& \partial_1 (M^a M^a)  =  \pm 2 \cos{\beta} \left [ M^a M^a -
\left (M^a \chi^a \right)^2 \right ],\nonumber \\    
&& \partial_1(M^a \chi^a )  =  \pm \sin{\beta} \left [ M^a M^a -
\left (M^a \chi^a \right)^2 \right ].
\label{eq4.6.1}
\eea
\noindent We have made the following rescaling of the field variables,
\be
M^a \rightarrow \frac{e}{\kappa} M^a, \ \
A_1^a \rightarrow \frac{e}{\kappa} A_1^a, x \rightarrow \frac{\kappa}{e^2},
\label{eq6.2}
\ee 
\noindent while deriving the equation (\ref{eq4.6.1}). Note that all
the field variables as well as the space coordinate are now dimensionless
quantity.

We now discuss the solutions
of (\ref{eq4.6.1}) for $\beta \neq \frac{m \pi}{2}$ and
$\beta = \frac{m \pi}{ 2}$ separately, where $m$ is any integer.\\
(a) $\beta \neq \frac{m \pi}{2}$ : Note that,
\be
M^a \chi^a =  \frac{1}{2} \tan{\beta} \left ( M^a M^a - b \right )
\label{eq4.7}
\ee
\noindent for $\beta \neq \frac{m \pi}{2}$, where 
$b$ is the integration constant. 
Using the relation (\ref{eq4.7}) in the first equation of (\ref{eq4.6.1}),
$M^a M^a$ can be determined completely,
\be
M^a M^a = \pm 2 \delta \cot^2{\beta} \tanh \left [ \delta \cos{\beta}
(x - x_0) \right ] + b + 2 \cot^2{\beta}, \ \
\delta = (1 + b \tan^2{\beta} )^{\frac{1}{2}},
\label{eq4.7.1}
\ee
\noindent where $x_0$ is an integration constant. The finite energy field
configurations demand that the integration
constant $b \geq - \cot^2{\beta}$.
The asymptotic values of $M^a M^a$ and $\chi^a M^a$ are,
\bea
&& (M^a M^a)_{+} = 2 (1 \pm \delta) \cot^2{\beta} + b, \ \ \
(M^a M^a)_{-} = 2 (1 \mp \delta) \cot^2{\beta} + b,\nonumber \\
&& (\chi^a M^a)_{+} = (1 \pm \delta) \cot{\beta}, \ \ \
(\chi^a M^a)_{-} = (1 \mp \delta) \cot{\beta},
\label{e4.7.1}
\eea
\noindent where a subscript `plus' or `minus' denotes the value of the
quantity inside the bracket at $x=\infty$
or $x=-\infty$, respectively. We will follow this notation throughout this
paper. The topological charge, the Noether charge and the energy are,
\be
Y_1= \pm \frac{4 e^2}{\kappa} \delta \cot^2 {\beta}, \ \ \ 
Z_1= \pm 4 \frac{e^2}{\kappa} \delta \cot{\beta}, \ \ \
E= \frac{4 e^2}{\kappa} \delta \frac{\cot{\beta}}{\sin{\beta}}.
\label{e4.7.2}
\ee
\noindent Note that $Y_1 = E \cos{\beta}$ and
$Z_1= E \sin{\beta}$. As a result, $E=\sqrt{Y_1^2 + Z_1^2}$, very much like
dyons in YMH theory. Also, note that
$Y_1$, $Z_1$ and $E$ are dependent on the
integration constant $b$ through $\delta$. For any value of
$\beta$, one
can make all these quantity to be zero by fixing $b$ to take its
minimum allowed value, i. e., $b=- \cot^2{\beta}$, or 
equivalently $\delta=0$.
This describes trivial vacuum solution. Finite energy nontrivial soliton
solutions
exist for $\delta > 0$.

We now choose consistently
$A_1^a$ to be zero. This implies that $\chi^a$ and $M^a$ are not independent,
\be
M^a = \eta_0^a + \cot{\beta} \chi^a,
\label{eq4.7.2}
\ee
\noindent where $\eta_0^a$ are three independent constants. Plugging back 
this expression into the second Bogomol'nyi equation and making use of Eq.
(\ref{eq4.7.1}), we determine $\chi^a$,
\be
\chi^a = \pm \frac{\eta_0^a}{\delta} \tan{\beta} \tanh{[\delta \cos{\beta}
(x - x_0)]} \pm \theta^a {\rm{sech}}[ \delta \cos{\beta} (x-x_0)].
\label{eq4.7.3}
\ee
\noindent The constants $\eta_0^a$ and $\theta^a$ satisfy the following
relations in order to maintain the unit norm of $\chi^a$,
\be
\theta^a \theta^a = 1, \ \ \theta^a \eta_0^a = 0, \ \
\eta_0^a \eta_0^a = \delta^2 \cot^2{\beta}.
\label{eq4.8.3}
\ee
\noindent 
One particular choice of $\theta$ and $\eta_0$ satisfying the relation
(\ref{eq4.8.3}) is $\theta=(0, 0, 1)$ and 
$\eta_0=( \delta \cos{\psi} \cot{\beta}, \delta \sin{\psi} \cot{\beta}, 0)$,
where $\psi$ is an arbitrary angle. In general, $\theta^a$ can be
parameterized as the
coordinates of unit sphere and $\eta_0^a$ as the coordinates of a sphere of
radius $\delta \cot{\beta}$. This reduces the number of constants to four.
This can further be reduced to three by imposing the condition $\theta^a
\eta_0^a =0$. 

(b) $\beta=\frac{m \pi}{2}$ :
We now discuss the special case $\beta=\frac{(2 m+1) \pi}{2}$.
All the components of $M^a$ are constant, $M^a= \xi^a$.
The quantity $\xi^a \chi^a$ is determined as,
\be
M^a \chi^a = \xi^a \chi^a = \mp p \tanh{[(x-x_0) p]},
\ \ p=\xi^a \xi^a.
\label{eq5.1}
\ee
\noindent Using this expression, we find,
\be
\chi^a = \mp \left [ \frac{\xi^a}{p} \tanh{[(x-x_0) p]} +
 \eta^a {\rm{sech}} [(x-x_0) p] \right ],
\label{eq5.2}
\ee
\noindent where $\eta^a$ are three different integration constants
having the properties,
\be
\eta^a \xi^a = 0, \ \ \eta^a \eta^a =1.
\label{eq5.3}
\ee
\noindent These properties of $\eta^a$ and $\xi^a$ are necessary in order to
maintain
the unit norm of $\chi^a$. 
Now note that $M^a \chi^a$
goes to $\mp p$ at $x=+\infty$, while it is $\pm 1$ at $x= - \infty$. Thus,
$Z_1 = \mp 2 p e$ and $Y_1 = 0$. This is the zero topological charge sector.
The purely topological sector is given by $\beta=m \pi$. In this case,
the role of $\chi^a$ gets exchanged with $M^a$.

\section{$U(1)$ Gauged Sigma Model}

The self-dual $U(1)$ gauged sigma model with pure CS dynamics in 2+1
dimensions is given by \cite{gg},
\be
{\cal{L}}_3 = \frac{1}{2} D_\mu \vec{\phi} . D^\mu \vec{\phi} + \frac{k}{4}
\epsilon^{\mu \nu \lambda} A_\mu F_{\nu \lambda} - \frac{1}{2 k^2}
\left (\phi_1^2 + \phi_2^2 \right ) \left ( v - \phi_3 \right )^2,
\label{eq0.0}
\ee
\noindent where $\vec{\phi}$ is a three component real scalar
field, $\vec{\phi} = \hat{n}_1 \phi_1 + \hat{n}_2 \phi_2 + \hat{n}_3 \phi_3$,
with unit norm in the internal space. The covariant derivative is defined as,
\be
D_\mu \vec{\phi} =
\partial_\mu \vec{\phi} + A_\mu \hat{n}_3 \times \vec{\phi},
\label{eq0.1}
\ee
\noindent and the field strength $F_{\mu \nu}= \partial_\mu A_\nu
- \partial_\nu A_\mu$. The scalar potential in (\ref{eq0.0}) has three
degenerate minima
for $0 \leq {\mid v \mid} < 1$. The symmetric phases are described by
$\phi_3=\pm
1$, while the asymmetric phase is at $\phi_3 = v$. These three minima merge
into two for ${\mid v \mid} \geq 1$ and we are left with 
only symmetric phases
of the theory.
The Lagrangian (\ref{eq0.0}) admits topological as well 
nontopological soliton
solutions with nonzero Noether charge.

We now obtain the 1+1 dimensional model corresponding to (\ref{eq0.0})
, following the same procedure as in the case of the completely gauged sigma
model, with the identification $A_2=N$,
\be
{\cal{L}}_4 = \frac{1}{2} D_\mu \vec{\phi} . D^\mu \vec{\phi} + \kappa N F_{01}
- \frac{1}{2} N^2 (\phi_1^2 + \phi_2^2) - \frac{1}{2 \kappa^2}
(\phi_1^2 + \phi_2^2) (v - \phi_3)^2, \ \ \ \mu=0, 1.
\label{eq1}
\ee
\noindent The potential in (\ref{eq1}) has three degenerate minima,
(i) $\phi_3=\pm 1$, $N \equiv $ arbitrary constant and 
(ii) $\phi_3 = v, N=0$
for $ 0 \leq {\mid v \mid} < 1$. For ${\mid v \mid} \geq 1$, there are
only two degenerate minima in the symmetric phase.

The equations of motion which follow from Eq. (\ref{eq1}) are,
\be
\kappa F_{01} = N (\phi_1^2 + \phi_2^2), \ \ \
\kappa N^\prime  =  j_0 ,\ \ \
\kappa \partial_0 N = j_1, \ \ \
j_\mu = \phi_2 D_\mu \phi_1 - \phi_1 D_\mu \phi_2,
\label{eq2.2}
\ee
\be
D_\mu \vec{J}^\mu  = - \left (\hat{n}_3 \times \vec{\phi} \right ) \left (
N^2 \phi_3 + \frac{1}{k^2} (v - \phi_3) ( v \phi_3 + 1 - 2 \phi_3^2) \right ).
\label{eq2.3}
\ee
\noindent
The current $\vec{J}_\mu =
\vec{\phi} \times D_\mu \vec{\phi}$ and the $U(1)$ current
is given by $ j_\mu = - \vec{J}_\mu . \hat{n}_3$. The Noether charge $q$ is
determined in terms of the asymptotic behaviour of $N$
from ({\ref{eq2.2}) as,
\be
q=\kappa \int dx N^\prime =  \kappa
[ N_{+} - N_{-}].
\label{eq}
\ee
\noindent Note that the nonzero Noether charge sectors are
characterized
by, $N_{+} \neq N_{-}$. 

We define two different conserved charges as follows,
\be
Y_2 = \frac{1}{2 \kappa} \int dx \left [ (v-\phi_3)^2 - \kappa^2 N^2 \right ]
^\prime, \ \ \ Z_2= \int dx \left [ N (v-\phi_3) \right ]^\prime.
\label{eq2.4}
\ee
\noindent We identify $Y_2$ and $Z_2$ as topological and Noether charge
, respectively. To see this, note that the momentum along the compactified
dimension,
$P_y= \int dx D_0 \vec{\phi} . D_2 \vec{\phi}$, can be expressed in terms
of the
asymptotic values of $N$ with the help of Gauss law. In particular, $P_y
= \frac{q}{2} (N_+ + N_-) = q \bar{N}$. We rewrite $Y_2$ in terms of
the asymptotic values of the field variables,
\be
Y_2 = \frac{1}{2 \kappa} \left [ (v - \phi_3)_+^2 - (v -\phi_3)_-^2
- 2 \kappa P_y \right ].
\label{eq2.5}
\ee  
\noindent Notice that $Z_2= \frac{v \mp 1}{\kappa} q$, in case,
$\phi_3$ interpolates from any one of the symmetric vacua to the
asymmetric vacuum.
This is also true when $\phi_3$ interpolates between the same symmetric
phase. However, $Z_2$ receives an extra 
contribution, $Z_2 = \frac{v}{\kappa}
q \mp 2 \frac{P_y}{q}$, in case it interpolates between the different
symmetric phases. In a broad sense, $Z_2$ can thus be regarded as the
Noether charge. 

The energy functional corresponding to (\ref{eq1}) is,
\be
E = \frac{1}{2} \int dx \left [ D_0 \vec{\phi} . D_0 \vec{\phi} +
D_1 \vec{\phi} . D_1 \vec{\phi} + N^2 ( \phi_1^2 + \phi_2^2 )
+ \frac{1}{\kappa^2} (\phi_1^2 + \phi_2^2 ) ( v -\phi_3 )^2 \right ].
\label{eq3.0}
\ee
\noindent The term $\kappa_0 N F_{01}$ do not contribute to the energy
functional, since it is first order in space-time derivative.
Let us now introduce two orthogonal vectors $\vec{A}$ and $\vec{B}$
as follows,
\be
\vec{B}= \hat{n}_3 \times \vec{\phi}, \ \ \vec{A}= \vec{\phi} \times \vec{B}.
\label{eq3.1}
\ee
\noindent These two vectors have the following properties,
\be
\vec{A}. \vec{A} = \vec{B} . \vec{B} = \phi_1^2 + \phi_2^2, \ \
\vec{A}. \vec{B} = 0, \ \ D_\mu \vec{\phi} . \vec{A} =
\partial_\mu \phi_3, \ \
D_\mu \vec{\phi} . \vec{B} = - j_\mu.
\label{eq3.2}
\ee
\noindent Using Eqs. (\ref{eq3.1}) and Eqs. (\ref{eq3.2}), the energy
functional
(\ref{eq3.0}) can be conveniently rewritten as,
\be
E = \frac{1}{2} \int dx \left [ \left ( D_0 \vec{\phi} \mp \vec{B}
P \right )^2  
+ \left ( D_1 \vec{\phi} \pm \vec{A} Q \right )^2 \right ]
\pm ( Y_2 \cos{\alpha} + Z_2 \sin{\alpha}),
\label{eq3.3}
\ee
\noindent where are $P$ and $Q$ are defined as,
\be
P= N \cos{\alpha} - \frac{1}{\kappa} (v-\phi_3) \sin{\alpha},\ \
Q= N \sin{\alpha} + \frac{1}{\kappa} (v -\phi_3) \cos{\alpha}.
\label{eq3.4}
\ee
\noindent Note that the lower bound on the energy functional (\ref{eq3.3}),
i.e. , $ E \geq {\mid Y \cos{\alpha} + Z \cos{\alpha} \mid}$, is expressed
as a linear combination of the Noether charge and the topological charge.
This is reminiscent
of what happens in the case of dyons in the 3+1 dimensional YMH theory.

The Bogomol'nyi bound is saturated, when the following first order equations
hold true,
\be
D_0 \vec{\phi} \mp \vec{B} P =0, \ \ D_1 \vec{\phi} \pm \vec{A} Q = 0.
\label{eq3.5}
\ee
\noindent These two first order equations are of course consistent with the
field equations (\ref{eq2.2}) and (\ref{eq2.3}). With the help
of the first
equation of (\ref{eq3.5}), the Gauss law can be conveniently rewritten as,
\be
\kappa N^\prime = \mp (\phi_1^2 + \phi_2^2) P.
\label{eq3.6}
\ee
\noindent Using the stereographic projection,
\be
u_1=\frac{\phi_1}{1+\phi_3}, \ \ u_2 = \frac{\phi_2}{1+\phi_3}, \ \ u = u_1 +
i u_2,
\label{eq3.7}
\ee
\noindent the second equation of (\ref{eq3.5}) is transformed as,
\be
(\partial_1 + A_1 ) u \mp Q u =0.
\label{eq3.8}
\ee
\noindent The gauge potential $A_1$ is determined in terms of 
the argument of $u$,
$A_1 = - [ Arg (u) ]^\prime$ and, hence can be consistently chosen as zero. 
We get 
the decoupled second order equation in terms of $\rho={\mid u \mid}^2$ after
combining
Eq. (\ref{eq3.6}) with (\ref{eq3.8}),
\be
\frac{\partial^2}{\partial x^2} ln \rho = \frac{\rho}{(1+\rho)^3} \left [
(v-1) + (v+1) \rho \right ].
\label{eq3.9}
\ee
\noindent We have scaled $x$ as $ x \rightarrow \frac{\kappa}{\sqrt{8}} x$
in the above
equation. Eq. (\ref{eq3.9}) is precisely the one dimensional version of the
decoupled
equation
obtained in the 2+1 dimensional $U(1)$ self-dual gauged sigma model
(\ref{eq0.0}). However, no exact solutions
is known in the 2+1 dimensional
case. Eq. (\ref{eq3.9}) can be written as a first order
nonlinear equation,
\be
\frac{\partial \rho}{\partial x} = \pm \frac{\rho}{1+\rho} \left [ a + b \rho
+ a_0 \rho^2
\right ]^{\frac{1}{2}},
\label{eq3.10}
\ee
\noindent where $a=a_0 - 2 v$, $b= 2 ( a_0 - v -1)$ and
$a_0$ is the integration
constant. Note that Eq. (\ref{eq3.10}) with the upper sign can be related
to the
same equation with the lower sign, by changing $x \rightarrow - x$. Hence,
we will consider the lower sign only now onwards. Once $\rho$ is known
from Eq. (\ref{eq3.10}), $N$ can be determined as,
\be
N = \frac{1}{\kappa} \left [ \frac{\sqrt{2}}{\sin{\alpha}}
\frac{\rho^\prime}{
\rho} - \cot{\alpha} \frac{(v-1)+(v+1) \rho}{1+\rho} \right ] .
\label{eq3.10.1}
\ee
\noindent We present some exact solutions of (\ref{eq3.10})
below.

No exact solution of (\ref{eq3.10}) is known for arbitrary $a_0$ and $v$.
We first try the simplest case $\delta_0= 4 a a_0 - b^2=0$. The
constants are determined as, $a_0=\frac{1}{2} (1+v)^2$, $a=\frac{1}{2}
(1-v)^2$
and $b=v^2-1$.
The problem now is to find the solution of the algebraic equation,
\be
\rho \left ( 1 + \frac{b}{2 a} \rho \right )^{\gamma - 1} =
\frac{2 a}{b^\gamma} e^{- \sqrt{a} (x-x_0)},
\label{eq3.11}
\ee
\noindent for $a, a_0 > 0$,
where $\gamma=\sqrt{\frac{a}{a_0}}$ and $x_0$ is the
integration constant. The most obvious choice now is to choose $\gamma=1$.
However, the solution $\rho= - e^{-\frac{1}{\sqrt{2}} (x-x_0)}$ is not
a physical one, because of the presence of a minus sign in front of it.
The next possibility, $\gamma=0$, is ruled out since 
$`a'$ is zero in this case. The algebraic equation (\ref{eq3.11}) certainly
can be
solved for $\gamma=\frac{1}{2}, \frac{1}{3}, \frac{1}{4}, \frac{2}{3},
\frac{3}{4}$ and their reciprocals. This is because Eq. (\ref{eq3.11})
is at most quartic in $\rho$ for these values of $\gamma$. 
It may be possible
to solve (\ref{eq3.11}) for other values of $\gamma$ also. 
However, there is no
general procedure for finding roots of fifth or higher order polynomial
equations and we do not discuss such cases here.
Now note that for a fixed $\gamma$, $v$ is completely determined. In
fact, for each value of $\gamma$, $v$ has two different values,
\be
v = - \frac{\gamma - 1}{\gamma+1}, \ \ \ - \frac{\gamma+1}{\gamma-1} .
\label{3.11.1}
\ee
\noindent These two values of $v$ are reciprocal to each other.
Consequently, for a fixed value of $\gamma$, once we know a solution
at a particular
$v$, we also know the solution at $\frac{1}{v}$. Also, note that $\gamma
\rightarrow \frac{1}{\gamma}$ implies $v \rightarrow - v$.

We now present some exact soliton solutions and their asymptotic behaviour
for different values of $\gamma$,\\
(i) For $\gamma= \frac{1}{2}$, we have the following two solutions in
terms of the variable $X=\sqrt{a} (x-x_0)$,
\be
\rho_{1,2} = a e^{- X} \left [ e^{-X} \pm \left ( e^{- 2 X} + 4 \right )^
{\frac{1}{2}} \right ], 
\label{eq3.12}
\ee
\noindent where $\rho_{1}(\rho_2)$ denotes the solution with the upper(lower)
sign. The field variable $N$ is not nonsingular 
all over the real line for $\rho_2$ and, hence is not a finite energy
solutions.
On the other hand, $\rho_1$ vanishes at $x=\infty$ and diverges at $x=
- \infty$. This implies that $\phi_3$ interpolates from $-1$ to $1$.
The asymptotics of $N$ for this solution is,
\be
N_+ = \frac{1}{\kappa} \left [ - \frac{\sqrt{2 a}}{\sin{\alpha}}
- (v-1) \cot{\alpha} \right ], \ \ \
N_- = \frac{1}{\kappa} \left [ - \frac{ 2 \sqrt{2 a}}{\sin{\alpha}}
- (v+1) \cot{\alpha} \right ] .
\label{eq3.12.1}
\ee
\noindent This solution $\rho_2$ is valid for both $v=\frac{1}{3}$ and
$v=3$.\\
(ii) We have the following solutions for $\gamma=2$, 
\be
\rho_{1,2} = \frac{a}{b} \left [ -1 \pm \left (1+ \frac{4}{b} e^{- X} 
\right )^{\frac{1}{2}} \right ].
\label{3.13}
\ee
\noindent This solution is valid for both $v=-\frac{1}{3}$ and $v=-3$. 
However,
$b$ is negative for $v=-\frac{1}{3}$ and $\rho_{1,2}$ becomes imaginary
for certain values of $x$. Thus, solutions with $v=-\frac{1}{3}$
can not be physical. On the other hand,
$\rho_2$ goes to $- \frac{2 a}{b}$ as $x \rightarrow \infty$. This means
$\phi_3$ goes to $v=-3$ at spatial infinity. Unfortunately, $\phi_3$
can not take this value because of the constraint 
$\vec{\phi} . \vec{\phi} =1$.
Thus, the only acceptable solution is $\rho_1$ with $v=-3$, which vanishes
at one end and diverges in the other end, 
implying that $\phi_3$ interpolates
between the symmetric vacua.
The behaviour of $N$ corresponding to this solution
is,
\be
N_+= \frac{4}{\kappa} \left [\cot{\alpha} -
\frac{1}{\sin{\alpha}} \right ], \ \ \
N_- = - \frac{1}{\kappa \sin{\alpha}} + \frac{2}{\kappa} \cot{\alpha} .
\label{eq3.13.a}
\ee
\noindent 
(iii) For $\gamma=\frac{3}{2}$ and  $v=-\frac{1}{5}$, we determine $\rho$
as,
\be
\rho = \frac{1}{2} + \frac{1}{4} B^{\frac{1}{3}} + B^{-\frac{1}{3}}, \ \
B= \frac{1}{2} \left [ 16 + 225 e^{- 2 X} + 15 e^{-X} \left ( 
32 + 225 e^{-2 X} \right )^{\frac{1}{2}} \right ] .
\label{eq3.13.b}
\ee
\noindent As $x \rightarrow \infty$, $\rho \rightarrow \frac{3}{2}$ and 
$\rho$ diverges as $x \rightarrow -\infty$. This is the 
solution interpolating
between the symmetric and the asymmetric vacuum. In particular, $\phi_3$
interpolates from $-1$ at $x=-\infty$ to $v=-\frac{1}{5}$ at $x=\infty$.
The behaviour of $N$ corresponding to this solution is,
\be
N_+= 0, \ \ \
N_- = -\frac{4}{5 \kappa \sin{\alpha}} - \frac{4}{5 \kappa} \cot{\alpha} .
\label{eq3.13.c}
\ee
\noindent We find exact solutions
for (a) $\gamma=\frac{1}{3}, v=2$, (b) $\gamma=3, v=-2$,
(c) $\gamma=\frac{2}{3}, v=\frac{1}{5}$,
(d) $\gamma=\frac{1}{4}, v=\frac{5}{3}$ and
(e) $\gamma=\frac{3}{4}, v=7$. All of these solutions interpolate
from $-1$ to $1$ and we present some of these solutions in Appendix A.
The solutions for other values of $\gamma$ and $v$ are not physical in the
sense that either they are of infinite energy or they become imaginary over
certain region of space.

Let us now consider the case $\delta_0 \neq 0$. The problem again
is to solve an algebraic equation similar to (\ref{eq3.11}), but
more complicated. We are able to solve this equation only for $v = 0$.
For this choice of
$v$,
$a=a_0$ and $b=2 ( a_0 -1)$. We have the following expression
for $\rho$ with $a_0 > 0$,
\be
\rho_{1,2} = \frac{1}{2 a_0} \left [ A \pm \left ( A^2 -
 4 a_0^2 \right )^{\frac{1}{2}} \right ], \ \
A = 2 a_0 \cosh^2{\frac{X}{2}} + (2 a_0 +1 ) \sinh^2{\frac{X}{2}}.
\label{eq3.14}
\ee
\noindent $\phi_3$ goes to $-1$ at both the spatial infinities for the
solution (\ref{eq3.14}). However, $N$ is not well behaved all over the
space for $\rho_2$. We discard this solution. The asymptotic behaviour
of $N$ corresponding to $\rho_1$ is given by,
\be
N_+ = \frac{1}{\kappa} \left [ \frac{\sqrt{2}}{\sin{\alpha}} -
\cot{\alpha} \right ], \ \ \
N_- = \frac{1}{\kappa} \left [ - \frac{\sqrt{2}}{\sin{\alpha}}
- \cot{\alpha} \right ] .
\label{eq3.14.a}
\ee
\noindent Note that the energy is expressed in terms of the topological
and the Noether charge as, $E=\sqrt{Y_2^2 + Z_2^2}$, for
all of these solutions. This is exactly like the energy bound in BPS dyons.
These solitons are domain walls in nature, interpolating between 
different symmetric and asymmetric vacua. 

\section{Summary and Discussions}

In conclusion, we have studied soliton solutions in certain
1+1 dimensional gauged sigma models. These models are 
obtained by dimensionally reducing 2+1 dimensional self-dual gauged
sigma models with pure CS dynamics. We have found a remarkable similarity
between these $1+1$ dimensional models and the $3+1$ dimensional YMH theory.
In particular, the Bogomol'nyi bound is expressed 
in terms of the topological
and the Noether charge in a similar way to that of the BPS dyons. Moreover,
the scale invariant solitons
with vanishing Noether charge, of the 2+1 dimensional completely gauged
sigma model, have definite scale and nonzero Noether charge in the
corresponding 1+1 dimensional theory. This resembles the way BPS
dyons can be obtained from four dimensional Eucledian Yang-Mills
theory. Such a similarity between Q-kinks
and the BPS dyons already exists. However, Q-kinks are necessarily time
dependent
solutions, while BPS dyons are static, minimum energy solutions of the YMH
theory. In our
case, the similarity is between the static solitons of gauged
sigma model with BPS dyons. Recently, it has been 
shown that the static solitons of dimensionally reduced self-dual $U(1)$
CS Higgs theory also share similar properties with the BPS dyons \cite{kor}.
However, the soliton solutions of the self-dual CS theory in 2+1 dimensions
are not scale invariant. Thus, the soliton solutions of completely
gauged sigma models studied in this paper  
have more similarities with the BPS dyons than any other existing models. 
Finally, we have obtained all static, exact soliton
solutions of the completely gauged sigma model saturating the Bogomol'nyi
bound. On the other hand,
we found only a class of exact, static soliton 
solutions for the $U(1)$ gauged
sigma model. The soliton solutions of both of these models are domain walls
in nature interpolating between different symmetric and asymmetric vacua. 

The models considered in this paper have no kinetic energy term corresponding
to the gauge fields and the gauge field equations appear as constraints. This
is because they are dimensionally reduced version of $2+1$ dimensional models
with pure CS dynamics. In this regard, one might also consider the 
dimensionally reduced version of
the $2+1$ dimensional completely gauged sigma model with both
Yang-Mills as well
as CS dynamics \cite{g1}. The resulting Lagrangian would have not only the
gauge field kinetic energy term, but also a kinetic 
energy term for the triplet
$M$ in terms of its covariant derivative and certain interaction term
dictated
by the $2+1$ dimensional anomalous magnetic moment 
interaction term. We expect
that all the results obtained in this paper will go through in a
straightforward way for this case also. 

It is known that gauged sigma models can be viewed as
a low energy effective action of certain gauged linear sigma models
\cite{dur}. These linear  
models are useful in studying different kinds of $2+1$ dimensional soliton
solutions in an unified manner. However,
in general, it is difficult to analyze the Bogomol'nyi equations arising out
of these linear sigma models in detail. 
The study of $1+1$ dimensional version
of these models, which are expected to be the gauged linear sigma models
corresponding to the models studied in this paper, may shed some light on
the $2+1$ dimensional problem. Similar considerations also apply for
the gauged $CP^N$ models. 

\acknowledgements{I would like to thank Avinash Khare for a careful
reading of the manuscript and valuable comments.}

\begin{appendix}

\section{Some more exact Solutions}

In this appendix, we present some more exact solutions of Eq.
(\ref{eq3.11}) for different values of $\gamma$ and $v$. For all of
these solutions, $\phi_3$ interpolate between $-1$ and $+ 1$. Also,
the energy is expressed in terms of the topological and Noether charge
as, $E=(Y_2^2 + Z_2^2 )^{\frac{1}{2}}$.\\
(a) $\gamma=\frac{1}{3}$, $v=2$:\\
\bea
& & \rho= e^{- 3 X} + (\frac{2}{3 C})^{\frac{1}{3}} e^{- 3 X} \left (
2 + 3 e^{- 3 X} \right ) + (\frac{C}{18})^{\frac{1}{3}},\nonumber \\
&& C = 3 e^{-3 X} \left [ 1 + 6 e^{-3 X} + 6 e^{-6 X} + 
\left ( 1 + \frac{4}{3} e^{-3 X} \right )^{\frac{1}{2}} \right ].
\eea
(b) $\gamma=3$, $v=-2$:\\
\be
\rho=-2 + D^{\frac{1}{3}} + D^{-\frac{1}{3}}, \ \
D= \frac{1}{2} \left [ 2 + 3 e^{-X} + \sqrt{3} e^{- \frac{X}{2}} \left
( 4 + 3 e^{-X} \right )^{\frac{1}{2}} \right ].
\ee
(c) $\gamma=\frac{2}{3}$, $v=\frac{1}{5}$:\\
\be
\rho=\frac{2}{15} ({\frac{2}{E}})^{\frac{1}{3}} \left [
- 4 e^{- 3 X} + 2^{\frac{1}{3}} E^{\frac{2}{3}} \right ], \ \
E= e^{- 3 X} \left [ 15 + \left ( 225 + 32 e^{-3 X} \right )^{\frac{1}{2}}
\right ].
\ee
\end{appendix}

\end{document}